\documentclass[twocolumn,showpacs,preprintnumbers,amsmath,amssymb]{revtex4}
\usepackage{graphicx}
\usepackage{dcolumn}
\usepackage{bm}

\begin{document}

\title{The excitation operator approach to non-Markovian dynamics of quantum impurity models in the Kondo regime}
\author{Pei Wang}
\email{wangpei@zjut.edu.cn}
\affiliation{Institute of Applied Physics, Zhejiang University of Technology, Hangzhou, 310023, P. R. China}

\date{\today}

\begin{abstract}
We present a numerical method for studying the real time dynamics of a small interacting quantum system coupled to an infinite fermionic reservoir. By building an orthonormal basis in the operator space, we turn the Heisenberg equation of motion into a system of linear differential equations, which is then solved iteratively by constructing excitation operators. The application of our method depends on a layer structure in the operator space, which help us to turn an infinite linear system into a series of small systems. We apply the method to investigate the decoherence dynamics of quantum impurity models in the Kondo regime with a non-Markovian reservoir. Taking full account of environmental back-actions and electron-electron interactions, we find that the coexistence of the Kondo correlation and a non-Markovian reservoir induces coherence ringings, which will be suppressed by either driving the system away from the particle-hole symmetric point or changing the reservoir into a Markovian one.
\end{abstract}


\maketitle

\section{introduction}

The decoherence of a small quantum system coupled to a fermionic bath has recently attracted much attention~\cite{restrepo,paladino,grishin,sousa,segal,marquardt06,marquardt07,neder,lutchyn,yamada,tu08,zhang12,shi,tu11,lei,chen,marcos}, due to the fact that the fermionic bath manifests as an important source of decoherence in a wide range of electronic devices designed for solid-state quantum computers. In spite of considerable effort, a thorough understanding of the coherence dynamics of fermionic baths is still lack in the non-Markovian regime. The non-Markovian dynamics is difficult to address theoretically, because the traditional Born-Markov approximation is invalid when the relaxation time of the environment is comparably long and then the back-action of the environment plays an important role in the dynamics of the system. To fully take into account the back-actions, the system-environment coupling must be treated in a non-perturbative way. In recent years, several non-perturbative approaches have been suggested to derive the master equation in the existence of strong back-actions~\cite{marquardt06,tu08,zhang12,shi,tu11,lei,chen,marcos}, when the electron-electron interaction is absent or irrelevant to the non-Markovian dynamics.

The non-interacting models, however, fail to incorporate the physics in solid-state structures where the Coulomb interaction between electrons is greater than the electron kinetic energy. A well known paradigm is the Kondo effect, displayed in quantum dots in the Coulomb blockade regime. In the Kondo effect, the e-e interaction induces a strong correlation of electrons, which can only be understood from a many-particle point of view. Then it is obliged to study the interplay of correlation physics and non-Markovian dynamics. 

In this paper, we study the coherence dynamics of quantum dots in the Kondo regime coupled to a non-Markovian fermionic reservoir. The model is described by the Anderson impurity Hamiltonian, which can be written as
\begin{equation}\label{sysenvhamiltonian}
 \hat H = \hat H_S + \hat H_B + \hat H_V.
\end{equation}
Here $\hat H_S = \epsilon_d \sum_{\sigma} \hat c^\dag_{0\sigma} \hat c_{0\sigma} + U \hat c^\dag_{0\uparrow} \hat c_{0\uparrow} \hat c^\dag_{0\downarrow} \hat c_{0\downarrow}$ is the system Hamiltonian, where $\epsilon_d$ denotes the gate potential and $U$ the Coulomb repulsive interaction. And $\hat H_B= \sum_{k\sigma} \epsilon_k \hat c^\dag_{k\sigma} \hat c_{k\sigma}$ is the Hamiltonian for a non-Markovian fermionic reservoir, which is set with a finite bandwidth and a sharp edge. The coupling Hamiltonian is given by $\hat H_V = \sum_{k\sigma} V_{k} \left( \hat c^\dag_{0\sigma} \hat c_{k\sigma} + h.c.\right)$. 

To solve this problem, we develop the numerical excitation operator method on the basis of previous works by the author~\cite{pei12a,pei12b}. This method is designed for studying the real time dynamics of a strongly-correlated system driven out of equilibrium. As for quantum impurity models, it is distinguished from various approaches~\cite{schiro09,werner09,schmidt08,muehlbacher08,anders05,anders06,silva08,boulat08,meisner09,feiguin08,schoeller09,karrasch10,andergassen11,pletyukhov10,kennes12a,kennes12b,hackl07,hackl08,pei10} devised recently by the fact that both the Coulomb interaction and the system-environment coupling are dealt with in its full extent and at the same time the reservoir is set to be infinite. These are difficult to be fully realized in present approaches. 

The plan of the paper is the following. In Sec.~II we introduce the excitation operator method. Its application in quantum impurity models is demonstrated in Sec.~III. The results are discussed in Sec.~IV. Especially, we will discuss the intrinsic correlation between the non-Markovian dynamics and the Kondo physics. In Sec.~V, we discuss the suppression of the non-Markovian dynamics. We conclude with a summary and discussion of our method and results in Sec.~VI. 

\section{The excitation operator method}

The excitation operator method is designed for solving the Heisenberg equation of motion:
\begin{equation}
 \frac{d\hat O(t)}{dt} = i [\hat H, \hat O(t)],
\end{equation}
where $\hat O$ is the observable that we are interested in.

We choose an orthonormal basis $\{ \hat O_i \}$ in the operator space which contains all the linear operators mapping the Hilbert space into itself. Any two basis operators satisfy
\begin{equation}
 \langle \hat O_i,\hat O_j\rangle = \delta_{i,j},
\end{equation}
where the bracket denotes the inner product between two operators and is generally defined as
\begin{equation}
  \langle \hat O_i,\hat O_j\rangle \mathrel{\mathop:}= \frac{1}{\mathcal{N}} \textbf{Tr} [\hat O_i^\dag\hat O_j].
\end{equation}
Here $\mathcal{N}$ is the normalization factor. An arbitrary observable can be decomposed into the linear combination of the basis operators. Then our target is to solve the Heisenberg equations of the basis operators.  

The Heisenberg equation is solved by constructing the excitation operators $\hat A_i$ satisfying the eigen equations:
\begin{equation}\label{def_A}
 [\hat H,\hat A_i ] = \lambda_i \hat A_i,
\end{equation}
where $\hat H$ is the Hamiltonian of the system, and $\lambda_i$ the excitation energy of $\hat A_i$. We suppose that the excitation operators are expressed by the basis operators as
\begin{equation}\label{decom_A}
 \hat A_i = \sum_j \mathcal{A}_{i,j} \hat O_j.
\end{equation}
The coefficients matrix $\mathcal{A}$ needs to be determined. We then calculate the commutators between the Hamiltonian and the basis operators
\begin{equation}\label{commutator_O}
 [\hat H, \hat O_i]=\sum_j \mathcal{H}_{j,i} \hat O_j,
\end{equation}
and obtain a matrix $\mathcal{H}$. By substituting Eq.~\ref{decom_A} and~\ref{commutator_O} into Eq.~\ref{def_A}, we find that $\mathcal{A}$ is in fact the unitary transformation to diagonalize the matrix $\mathcal{H}$.

In principle, the elements of $\mathcal{H}$ can be written as
\begin{equation}
 \mathcal{H}_{i,j} = \langle \hat O_i, [\hat H,\hat O_j]\rangle.
\end{equation}
By using the definition of the inner product and the fact that the Hamiltonian is self-adjoint, we prove that $\mathcal{H}$ must be a Hermitian matrix and diagonalizable. 

The solution of the Heisenberg equation can be expressed by the coefficients $\mathcal{A}$ as
\begin{equation}
 \hat O_i (t)= \sum_{j,i'} \mathcal{A}^*_{j,i} e^{i\lambda_j t}\mathcal{A}_{j,i'} \hat O_{i'}.
\end{equation}
Here we use the fact that $\mathcal{A}$ is unitary. 

In practice, the dimension of $\mathcal{H}$ grows exponentially with the system size, so that directly diagonalizing it is impossible. However, there is a layer structure in the operator space, generated by the superoperator $[\hat H,\cdot]$. This indicates that we could change the problem of diagonalizing $\mathcal{H}$ into the problem of diagonalizing a series of small matrices.

We consider the evolution of the basis operator $\hat O_i$ in a small time interval $\tau$. The solution of the Heisenberg equation $\hat O_i(\tau)$ is mostly limited in a subspace of the whole operator space, generated by $\hat O_i$ and $[\hat H, \hat O_i]$. As $\tau \to 0$, we can calculate $\hat O_i(\tau)$ in this subspace, the dimension of which is small. In other words, we express $[\hat H, \hat O_i]$ as
\begin{equation}
 [\hat H, \hat O_i]=\sum_j \tilde {\mathcal{H}}_{j,i} \hat O_j,
\end{equation}
where $\tilde {\mathcal{H}}_{j,i}$ is non-zero as $j\neq i$. Obviously, $\tilde {\mathcal{H}}$ is a submatrix of $\mathcal{H}$. As $\tau\to 0$, the solution can be written as
\begin{equation}\label{intervalevolution}
 \lim_{\tau\to 0} \hat O_i (\tau) = \sum_{j,i'} \tilde{\mathcal{A}}^*_{j,i} e^{i \tilde{\lambda}_j \tau} \tilde{\mathcal{A}}_{j,i'} \hat O_{i'},
\end{equation}
where $\tilde {\lambda}_i$ and $\tilde{\mathcal{A}}$ are the eigenvalues and the unitary matrix of $\tilde {\mathcal{H}}$ respectively. 

To calculate $\hat O_i(t)$ at a finite time, we divide the time $t$ into $N$ small intervals of length $\tau=t/N$, and have
\begin{equation}
\begin{split}
& \hat O_i(t) \\ & = e^{i \hat H \tau }\left( e^{i \hat H \tau }  \left( \cdots \left( e^{i \hat H \tau} \hat O_i e^{-i\hat H  \tau} \right) \cdots \right) e^{-i \hat H \tau }\right) e^{-i \hat H\tau}.
\end{split}
\end{equation}
In each time interval, the evolution of the basis operators is calculated according to Eq.~\ref{intervalevolution}. As $\tau\to 0$, the result will go to the solution of the Heisenberg equation. The point of this method is to utilize the layer structure in the operator space. That is, the whole basis can be gradually generated from a single basis operator $\hat O_i$ by iteratively calculating the commutators between the Hamiltonian and the basis operators. But there is no such a structure in the Hilbert space. This is why we choose to solve the Heisenberg equation, instead of the Schr\"{o}dinger equation.

The number of basis operators which need to be stored in calculating $\hat O_i(N\tau)$ increases exponentially as $N$ increasing. This is a problem for numerical calculations. It can be solved by a truncation scheme: after obtaining $\hat O_i(N\tau)$ one keeps only the $M$ basis operators with the largest amplitudes. In this way, the number of the stored basis operators is fixed to be $M$ and the computation time increases linearly with $N$. Suitable values for the parameter $M$ depends upon the model. It should be decided numerically by varying $M$. 

\section{The model and the orthonormal basis of the operator space}

Our system consists of an impurity site, which is coupled, via particle-particle exchanges, to an electron reservoir. The system plus environment is described by the Hamiltonian~[\ref{sysenvhamiltonian}]. To facilitate applying the excitation operator method, we re-express the Hamiltonian of the reservoir in real space in terms of an infinite chain:
\begin{equation}
 \hat H_B = -g \sum_{\sigma,i=1}^\infty (\hat c_{i\sigma}^\dag \hat c_{i+1,\sigma}+h.c.) ,
\end{equation}
where $\hat c_{i\sigma}$ is the electron annihilation operator at site $i$. We take the size of the reservoir to be infinite. This avoids any coherence oscillation due to the finite-size effects. The coupling Hamiltonian now becomes
\begin{equation}
\begin{split}
 \hat H_V = V \sum_\sigma (\hat c_{0\sigma}^\dag\hat c_{1\sigma} + h.c.) ,
\end{split}
\end{equation}
where the system is coupled only to the first site of the reservoir. The system Hamiltonian keeps invariant as
\begin{equation}
 \hat H_S= \epsilon_d \sum_\sigma \hat c_{0\sigma}^\dag\hat c_{0\sigma} + U \hat n_{0 \uparrow} \hat n_{0\downarrow}.
\end{equation}

We find an orthonormal basis in the operator space of this model by transforming it into a spin-$\displaystyle\frac{3}{2}$ chain by the Jordan-Wigner transformation. 

The model contains a series of sites. The dimension of the local Hilbert space at each site is four with the basis vectors $|0\rangle$, $|\uparrow\rangle$, $|\downarrow\rangle$ and $|\uparrow \downarrow\rangle$. Keeping in mind that the basis operators are orthogonal to each other, we choose next sixteen $4\times 4$ matrices as the local basis operators: $\left( \begin{array}{cc} \mathbf{1} & 0 \\ 0 & \mathbf{1} \end{array} \right)$, $\left( \begin{array}{cc} \mathbf{1} & 0 \\ 0 & -\mathbf{1} \end{array} \right)$, $\left( \begin{array}{cc} o^\alpha & 0 \\ 0 & o^\alpha \end{array} \right)$, $\left( \begin{array}{cc} o^\alpha & 0 \\ 0 & -o^\alpha \end{array} \right)$, $\left( \begin{array}{cc} 0 & \mathbf{1} \\ \mathbf{1} & 0 \end{array} \right)$, $\left( \begin{array}{cc} 0 & -\mathbf{1} \\ \mathbf{1} & 0 \end{array} \right)$, $\left( \begin{array}{cc} 0 & o^\alpha \\ o^\alpha & 0 \end{array} \right)$ and $\left( \begin{array}{cc} 0 & -o^\alpha \\ o^\alpha & 0 \end{array} \right)$. Here the $\mathbf{1}$ denotes the two-dimensional identity matrix, and $o^\alpha$ with $\alpha=x,y,z$ the three generators of $SU(2)$ algebra, which are
\begin{eqnarray}
 o^x = \left( \begin{array}{cc} 0 & 1 \\ 1 & 0 \end{array}\right), o^y = \left( \begin{array}{cc} 0 & -1 \\ 1 & 0 \end{array}\right), o^z = \left( \begin{array}{cc} 1 & 0 \\ 0 & -1 \end{array}\right).
\end{eqnarray}
The sixteen matrices form a complete basis of the local operator space. 

A basis operator can be expressed as the tensor product of the local operators:
\begin{equation}\label{basisoperators}
 \hat O = \prod_{i=0}^\infty \otimes \sigma_i, 
\end{equation}
where $\sigma_i$ denotes the local operator at site $i$. Now we explicitly define the inner product as
\begin{equation}
  \langle \hat O_i,\hat O_j\rangle \mathrel{\mathop:}= \frac{1}{4^L} \textbf{Tr} [\hat O_i^\dag\hat O_j],
\end{equation}
where $L$ denotes the total number of sites and is taken as $L\to \infty$. It is easy to prove that the operators in Eq.~\ref{basisoperators} satisfy the orthogonal relations. 

The Hamiltonian is real, so that $\tilde{\mathcal{H}}$ is a real symmetric matrix. And the diagonal elements of $\tilde{\mathcal{H}}$ are all zero. Because the basis operators are either symmetric or anti-symmetric, then the projection of $[\hat H,\hat O_i]$ on $\hat O_i$ is zero.

The Hamiltonian can be expressed in the basis operators by the Jordan-Wigner transformation, in which a phase factor is attached to each site to produce the anti-commutative field operators. They are
\begin{eqnarray}\nonumber
 \hat c^\dag_{i\uparrow} &=& \prod_{j<i} \otimes \left( \begin{array}{cc} o^z & 0\\ 0 & - o^z \end{array}\right)_j \\ && \otimes \left( \frac{1}{2} \left( \begin{array}{cc} o^x & 0\\ 0 & o^x \end{array}\right)_i +  \frac{1}{2} \left( \begin{array}{cc} o^y & 0\\ 0 & o^y \end{array}\right)_i \right),
\end{eqnarray}
and
\begin{eqnarray}\nonumber
 \hat c^\dag_{i\downarrow} &=& \prod_{j<i} \otimes \left( \begin{array}{cc} o^z & 0\\ 0 & -o^z \end{array}\right)_j \\ && \otimes \left( \frac{1}{2}\left( \begin{array}{cc} 0 & -o^z\\ o^z & 0 \end{array}\right)_i +  \frac{1}{2}\left( \begin{array}{cc} 0 & o^z\\ o^z & 0 \end{array}\right)_i \right),
\end{eqnarray}
where $i,j=0,1,2,\cdots$ denote the sites and $\left( \begin{array}{cc} o^z & 0\\ 0 & -o^z \end{array}\right)_j$ the phase factor at site $j$.

The hopping term in the Hamiltonian can then be expressed as
\begin{eqnarray}\nonumber
&& \sum_\sigma \left( \hat c^\dag_{i\sigma}\hat c_{i+1,\sigma} + h.c.\right) \\ \nonumber && = \frac{1}{2} \left( \begin{array}{cc} o^x & 0 \\ 0 & -o^x \end{array}\right)_i \otimes \left( \begin{array}{cc} o^x & 0 \\ 0 & o^x \end{array}\right)_{i+1} \\ \nonumber && - \frac{1}{2} \left( \begin{array}{cc} o^y & 0 \\ 0 & -o^y \end{array}\right)_i \otimes \left( \begin{array}{cc} o^y & 0 \\ 0 & o^y \end{array}\right)_{i+1} \\ \nonumber && + \frac{1}{2} \left( \begin{array}{cc} 0 & 1 \\ 1 & 0 \end{array}\right)_i \otimes \left( \begin{array}{cc} 0 & o^z \\ o^z & 0 \end{array}\right)_{i+1} \\ && -\frac{1}{2} \left( \begin{array}{cc} 0 & -1 \\ 1 & 0 \end{array}\right)_i \otimes \left( \begin{array}{cc} 0 & -o^z \\ o^z & 0 \end{array}\right)_{i+1}.
\end{eqnarray}
And the system Hamiltonian becomes
\begin{eqnarray}\nonumber
 \hat H_S &=& \epsilon_d +\frac{U}{4} - \frac{2\epsilon_d +U}{4} \left( \begin{array}{cc}1 & 0 \\ 0 & -1 \end{array} \right)_0 \\ \nonumber && -  \frac{2\epsilon_d +U}{4}  \left( \begin{array}{cc} o^z & 0 \\ 0 & o^z \end{array} \right)_0+ \frac{U}{4}  \left( \begin{array}{cc} o^z & 0 \\ 0 & -o^z \end{array} \right)_0 . \\ 
\end{eqnarray} 

After solving the Heisenberg equation, we need to calculate the expectation value of the basis operators with respect to the initial state. This is done by transforming the basis operators into Majorana operators defined as
\begin{equation}
\hat \gamma_i^{\sigma \pm}= \hat c^\dag_{i\sigma} \pm \hat c_{i\sigma}.
\end{equation}
The sixteen local operators at site $i$ are in one-to-one correspondence with next Majorana operators: $\mathbf{1}$, $\gamma_i^{\uparrow +}$, $\gamma_i^{\uparrow -}$, $\gamma_i^{\downarrow +}$, $\gamma_i^{\downarrow -}$, $\gamma_i^{\uparrow +} \gamma_i^{\uparrow -}$, $\gamma_i^{\uparrow +} \gamma_i^{\downarrow +}$, $\gamma_i^{\uparrow +} \gamma_i^{\downarrow -}$, $\gamma_i^{\uparrow -}\gamma_i^{\downarrow +}$, $\gamma_i^{\uparrow -}\gamma_i^{\downarrow -}$, $\gamma_i^{\downarrow +} \gamma_i^{\downarrow -}$, $\gamma_i^{\uparrow +} \gamma_i^{\uparrow -}\gamma_i^{\downarrow +}$, $\gamma_i^{\uparrow +} \gamma_i^{\uparrow -} \gamma_i^{\downarrow -}$, $\gamma_i^{\uparrow +} \gamma_i^{\downarrow +}\gamma_i^{\downarrow -}$, $\gamma_i^{\uparrow -}\gamma_i^{\downarrow +}\gamma_i^{\downarrow -}$ and $\gamma_i^{\uparrow +}\gamma_i^{\uparrow -}\gamma_i^{\downarrow +}\gamma_i^{\downarrow -}$. They are not exactly the same, since the product of an odd number of Majorana operators, such as $\gamma_i^{\downarrow +}$, contains phase factors at the sites $j<i$. However, we can design an iterative algorithm to transform a basis operator into a product of Majorana operators. The algorithm begins from the largest site where the local operator is not the identity, and sweeps the chain in the descending order. 

After the transformation, the expectation value is calculated by using the Wick's theorem. The contraction of a pair of Majorana operators at zero temperature is found to be
\begin{eqnarray}
\langle \gamma_i^{\sigma +} \gamma_j^{\sigma +} \rangle = -\langle \gamma_i^{\sigma -} \gamma_j^{\sigma -} \rangle = \delta_{i,j},
\end{eqnarray}
and
\begin{eqnarray}
\langle \gamma_i^{\sigma +} \gamma_j^{\sigma -} \rangle = \frac{-2\sin(|i-j|\pi/2)}{|i-j|\pi},
\end{eqnarray}
as $|i-j|$ is an odd number.

\section{Interaction-induced coherence ringing in a non-Markovian environment}

We study the coherence dynamics of the system after its coupling to the reservoir is switched on at the time $t=0$. The reduced density matrix of the system is obtained by calculating the expectation values of the sixteen local operators, and is formally written as $ \sum_{i,j} \rho_{ij} |i\rangle \langle j|$, where $i,j=1,2,3,4$ and the corresponding states are $|0\rangle, |\uparrow\rangle , |\downarrow\rangle$ and $|\uparrow\downarrow\rangle$ respectively. Our method is distinguished from the master equation approach by the fact that no approximation is made on solving the Heisenberg equation and the environmental back-actions are fully taken into account. 

We set the reservoir at zero temperature, avoiding the thermal fluctuation which would suppress the Kondo resonance. The Fermi energy of the reservoir is set to be the energy zero. We employ the level-broadening at the impurity site $\Gamma$, generally defined as $\Gamma=V^2/g$~\cite{meisner09}, as the energy unit. This is usually in studying the Anderson impurity model. And the time unit is set to be $1/\Gamma$ (the convention $\hbar=1$ is used throughout the paper). 

At the particle-hole symmetric point, i.e., $\epsilon_d=-U/2$, a large $U$ provides a limit to the electron number of the system. The system is in the Kondo regime and can be described by a single spin. We suppose that its initial state is prepared as a superposition of the spin up and down states, i.e., $\alpha |\uparrow\rangle + \beta |\downarrow\rangle $. In the decoherence theory considering a Markovian environment, the coherence of the initial state will be lost in an exponential way after coupled to the reservoir. However, the real environment in the experiments is usually not Markovian, and the back-actions from the environment to the spin cannot be neglected. Here we consider a non-Markovian reservoir by setting the bandwidth of the reservoir to be comparable with the level-broadening at the impurity, i.e., $g \sim \Gamma$. 

\begin{figure}\label{}
\includegraphics[width=0.4\textwidth]{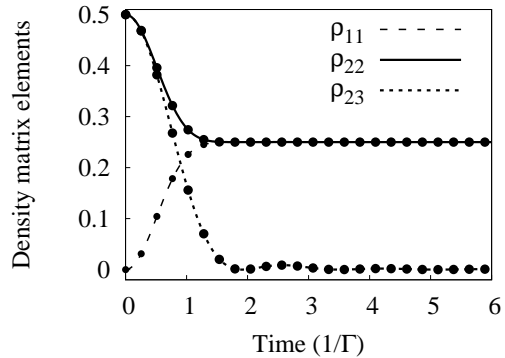}
\caption{Time evolution of the elements of the reduced density matrix at the symmetric point as $U=0$. The results by the excitation operator method, represented by the black circles, are compared with the exact solution, represented by the various types of lines. The initial state is set to be $\frac{1}{\sqrt{2}}\left( |\uparrow \rangle + |\downarrow\rangle \right)$. The coupling in the reservoir is set to be $g=\Gamma$. }
\end{figure}
We first set the interaction $U$ to zero to compare our result with the exact solution, obtained by exact diagonalization of the single-particle eigenmodes. The elements of the reduced density matrix are shown in Fig.~1. The result by the excitation operator method fits well with the exact solution, until the density matrix has relaxed to its equilibrium value. This proves that our method is a powerful tool in studying the real time dynamics of a quantum system coupled to a non-interacting reservoir. The errors can be controlled by letting $\tau \to 0$ and $M\to \infty$. It provides a reliable way of understanding the dynamics of decoherence, especially in strongly-correlated systems, where no analytical method is available. 

\begin{figure}\label{}
\includegraphics[width=0.4\textwidth]{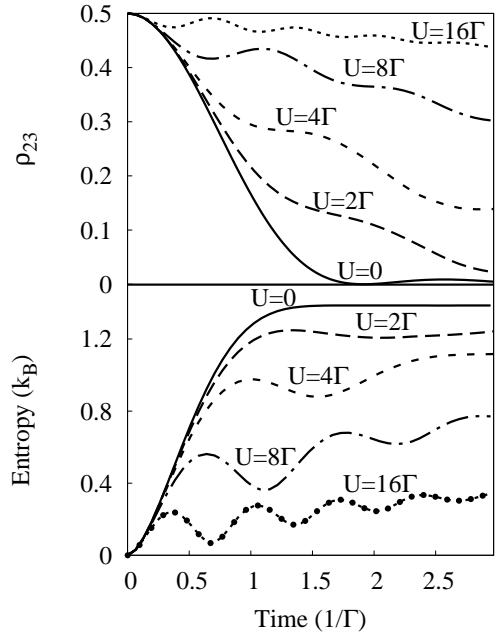}
\caption{The off-diagonal element and the von Neumann entropy of the density matrix as a function of time at different $U$ (top: off-diagonal element, bottom: entropy). We choose the initial state of the system to be $\frac{1}{\sqrt{2}} |\uparrow\rangle + \frac{1}{\sqrt{2}} |\downarrow\rangle$. The entropy for a different state $\sqrt{\frac{2}{3}} |\uparrow\rangle +  \sqrt{\frac{1}{3}} |\downarrow\rangle$ is also studied at $U=16\Gamma$, which is represented by the black circles for a comparison.}
\end{figure}
As the e-e interaction is absent, the off-diagonal element, i.e., the coefficient of the term $|\uparrow\rangle\langle \downarrow|$, decays exponentially, as predicted by the decoherence theory. This is the feature of a Markovian dynamics. The back-actions of the environment are sufficiently suppressed. However, it is not the case as $U\gg \Gamma$ (see the top panel in Fig.~2). In the existence of a strong interaction, the exponential decays are replaced by oscillations. And the intermediate quasi-steady regimes are observed. The decoherence time significantly increases as $U$ increasing. We then analyze the time evolution of the von Neumann entropy at different $U$ (see the bottom panel in Fig.~2). As is well known, the equilibrium value of the entropy is $2\ln 2$~\cite{bulla}. As $U=0$, the entropy increases monotonically from zero towards its equilibrium value, corresponding to the exponential decay of the off-diagonal elements. But as $U\gg \Gamma$, we find strong oscillations in the entropy, which is the signal of non-Markovian dynamics. The coherence in the initial state is lost and recovered repeatedly, similar to the spin echo effect. However, in our model, the purification of states arises naturally from the e-e interaction and no external driving field is needed as in the spin echo or dynamical decoupling technique~\cite{viola98}. This provides a new perspective in protecting the quantum state. 

The coherence ringing is an effect induced by the e-e interactions. It must be distinguished from the oscillations of coherence observed in the non-Markovian environments before~\cite{tu08,zhang12}, where the e-e interaction is absent. Without interactions, the electrons move independently, and the non-Markovian dynamics can be understood in the single particle picture. However, in the existence of a strong interaction, the single-particle picture breaks down due to the correlations between electrons. As $U\gg \Gamma$, the dissipation process is controlled by the Kondo correlation. The steady state as the time goes to infinity is a spin singlet. The correlation between the spin in the system and the spins in the reservoir is built in course of time, accompanied by the loss of coherence in the system. The non-Markovian coherence dynamics is in fact related to the dynamics of spin correlations in a Kondo model.

The initial state is found to be indifferent to the dynamics of decoherence. As an example, we choose two different initial states, which are $\frac{1}{\sqrt{2}} |\uparrow\rangle + \frac{1}{\sqrt{2}} |\downarrow\rangle$ and $\sqrt{\frac{2}{3}} |\uparrow\rangle +  \sqrt{\frac{1}{3}} |\downarrow\rangle$, and plot the corresponding entropies as a function of time in Fig.~2. The results are exactly the same. This reflects the spin-flip symmetry in the Hamiltonian. The correlation between the non-Markovian dynamics and the Kondo physics is universal for the initial state $\alpha |\uparrow\rangle + \beta |\downarrow\rangle$.

\begin{figure}\label{}
\includegraphics[width=0.4\textwidth]{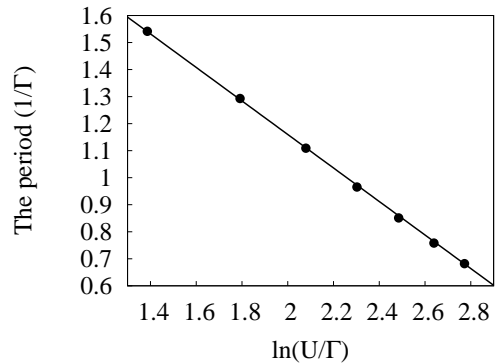}
\caption{In this figure, we plot the period of coherence ringing obtained from the numerical result, represented by black circles, as a function of $U$. The data is fitted to a linear function, represented by the solid line.}
\end{figure}
The period of the coherence ringing is obtained from the numerical result, and plotted as a function of $U$ in Fig.~3. In a large regime of $U$, the oscillation period is found to be proportional to $\ln\left(U/\Gamma \right)$. As the interaction strength increasing, the period decreases in a logarithmic way. That the oscillation period depends on $U$ can be understood by studying the energy levels of the system. We emphasize that the coherence ringing happens as the system is at the particle-hole symmetric point, i.e., $\epsilon_d = -U/2$. At the symmetric point, the Hamiltonian of the system changes into
\begin{eqnarray}
 \hat H_S =  \frac{U}{4}  \left( \begin{array}{cccc} 1 & 0 &0 & 0 \\ 0 & -1 & 0 & 0 \\ 0 & 0 & -1 &0 \\ 0 & 0 & 0 & 1 \end{array} \right) . 
\end{eqnarray}
We see that the energy levels of the system are degenerate. The ground level is two-fold degenerate, containing the spin up and down states. It is separated by a gap of $U/2$ from the excited level, which is also two-fold degenerate, containing the vaccum state and the doubly-occupied state. An energy gap of $U/2$ protects the sub-Hilbert space containing the states $|\uparrow\rangle$ and $|\downarrow\rangle$, and then protects the quantum coherence in the initial state. The gap is critical to the appearance of coherence ringing, which is obvious only as the gap is large. 

\section{suppression of the coherence ringing}

\begin{figure}\label{}
\includegraphics[width=0.4\textwidth]{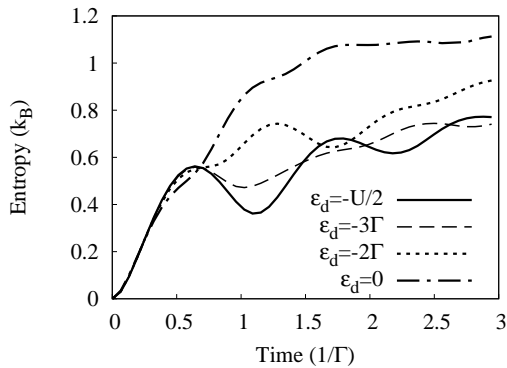}
\caption{The time evolution of the entropy at different $\epsilon_d$. The interaction is set to be $U=8\Gamma$. Then the particle-hole symmetric point is at $\epsilon_d=-4\Gamma$. }
\end{figure}
We attribute the coherence ringing to the coexistence of the Kondo correlation and the non-Markovian reservoir. Then it should disappear if any of the two conditions is broken. This is verified by the numerical results (see Fig.~4 and~5). 

In Fig.~4, we plot the time evolution of the entropy at different gate potentials. As the system is away from the particle-hole symmetric point, the coherence ringing is suppressed. This is due to the suppression of the Kondo resonance as the system is depleted or doubly-occupied. The suppression of coherence ringing can also be understood by the splitting of the excited level. The coherence ringing is distinguished from a simple Rabi oscillation because there are totally four levels in the system. By driving the system away from the symmetric point, we break the degeneracy at the excited level, which splits into the vacuum level and the doubly-occupied level. This is related to the disappear of the coherence ringing. According to our knowledge, it is the first time to find that the coherence in the ground state depends on the degeneracy at the excited level.

\begin{figure}\label{}
\includegraphics[width=0.4\textwidth]{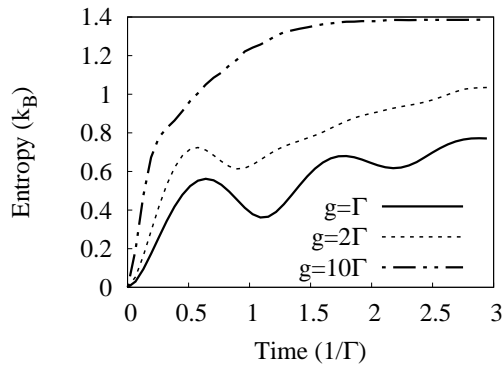}
\caption{The time evolution of the entropy at the particle-hole symmetric point at different $g$. The interaction is set to be $U=8\Gamma$. The bandwidth of the reservoir $g\to \infty$ corresponds to the Markovian limit. }
\end{figure}
The e-e interaction induces a coherence ringing only if the environment is non-Markovian, i.e., $g\sim \Gamma$. In Fig.~5, we show the entropy functions at different $g$, the bandwidth of the reservoir. In the case of $g=\Gamma$, the non-Markovian dynamics is significant. The coupling between the neighbor sites in the reservoir is as same as the coupling between the system and the reservoir. The back-action is strong since the relaxation time in the system is comparable with that in the reservoir. As the interaction is much larger than the bandwidth, i.e., $U\gg g$, the coherence ringing appears. If we keep the interaction invariant, at the same time increasing the bandwidth $g$, the coherence ringing is suppressed. As $g\sim U \gg \Gamma$, the oscillation is totally destroyed. In the limit of an infinite band, i.e., the Markovian limit, the entropy function recover the feature in the non-interacting model. That is, it increases monotonically towards the steady value: $2\ln 2$. But the relaxation time is now controlled by the interaction $U$, instead of the impurity level-width $\Gamma$. We see the disappear of the coherence ringing as the reservoir changes gradually to the Markovian limit.

\section{conclusions}

We have presented the numerical excitation operator method to coherence dynamics of an interacting quantum system coupled to a fermionic reservoir. Compared with the present analytical approaches, it takes full account of the Coulomb interaction between electrons and the system-environment coupling, and then provides new information on the interplay of electron-electron correlations and environmental back-actions. At the same time, our method takes into account an infinite reservoir by utilizing the layer structure in the operator space, and then avoids the finite-size effects which threaten the present numerical methods. We have applied this method to an interacting quantum dot coupled to a fermionic reservoir, discovering the coherence ringing induced by the e-e interaction in the Kondo regime. The coherence ringing is a many-body effect and can only be observed in the presence of both Kondo resonance and non-Markovian reservoir. It will be suppressed as the system is away from the particle-hole symmetric point or the reservoir changes into the Markovian limit.

Although we concentrate in this paper the dynamics of decoherence in quantum impurity models. The method that we presented can be applied to investigate the real time dynamics in a wide range of models describing a quantum system coupled to spin, fermionic and bosonic reservoirs. 

\section*{Acknowledgement}

I thank X. Wan, J. L. Wu, J. H. An and D. Suter for useful discussions.

\end{document}